%% file: main.tex
\documentclass[11pt,reqno]{amsart}
\usepackage{mypub}
\usepackage[colorlinks,citecolor=blue]{hyperref}
\usepackage{subfiles}
\begin{document}

\title[Non--isentropic Relativistic Euler]{The Non--isentropic Relativistic 
Euler System Written in a Symmetric Hyperbolic Form} 
  \author[U.~Brauer]{Uwe Brauer}
  \address{
    Uwe Brauer Departamento de Matemática Aplicada\\ Universidad
    Complutense Madrid 28040 Madrid, Spain} \email{oub@mat.ucm.es}
  \thanks{U.~B. 
    gratefully acknowledges support from Grant MTM2016-75465 by
    MINECO, Spain and UCM-GR17-920894.} 
\author[L.~Karp]{Lavi Karp}
  \address{
    Lavi Karp
    Department of Mathematics\\ ORT Braude College\\
    P.O. 
    Box 78, 21982 Karmiel\\ Israel}
  \email{karp@braude.ac.il}
\dedicatory{This paper is dedicated to our friend Michael Reissig}
\begin{abstract}
  We cast the non--isentropic relativistic Euler system into a symmetric
  hyperbolic form. 
  Such systems are very suited to treat initial value problems of
  hyperbolic type. 
  We obtain this form by using the pressure $p$ and not the density
  $\rho$ as a variable. 
  However, the system becomes degenerate when the pressure $p$
  approaches zero, and in these cases we regularise the system by
  replacing the pressure with an appropriate new matter variable, the
  Makino variable.
\end{abstract}

\maketitle{}

\subfile{sec-01-introduction.tex}

\subfile{sec-02-non-isentropic.tex}

\subfile{sec-03-system-p-u-S.tex}

\subfile{sec-04-makino.tex}

\bibliographystyle{amsalpha-url} 
\bibliography{bibgraf}

\end{document}

%% file: sec-01-introduction.tex
\section{Introduction}
\label{sec:introduction}

Existence and uniqueness theorems of a class of solutions have been
proved for the non--relativistic compressible Euler
equations for the isentropic case by 
\cite{makino_86
}, and later for the non-isentropic case by 
\cite{makino_ukai_kawashima_86
}. 

The situation, however, for the relativistic compressible Euler
equations is more involved.
The equivalent to the result obtained by Makino
\cite{makino_86
}, has been proven, for a restricted setting by Rendall,
\cite{Rendall92:_fluid
}, which was later extended by the authors 
\cite{BK8
} and \cite{BK3
}.

All those results had been obtained by casting, in one way or the other, the
Euler equations into a symmetric-hyperbolic first-order system. 
Such systems had been introduced Friedrich in 1954
\cite{Friedrichs_54}, and has been one of the most effective approaches
to prove the well--posedness (existence, uniqueness, and continuity of
the flow map) for these systems.

The non-isentropic case is more complicated.
Speck \cite{Speck_09
} studied the Cauchy problem for the Nordstr\"om scalar gravitational
field equation coupled to the non--isentropic Euler equations. 
He proved local existence, uniqueness and the continuity of the flow
map, but since he claimed that the system could not be cast into
symmetric hyperbolic form, he used Christodoulou's theory of the
energy current
\cite{christodoulou00A
} to obtain his results.

Choquet-Bruhat studied the Cauchy problem for both, the isentropic and
the non--isentropic, Einstein--Euler system, using Leray hyperbolic
systems
\cite{Foures-Bruhat_58
}.
Moreover, she also used a different method relying upon Leray-Ohya
hyperbolic systems, see 
\cite{choquet-bruhat66:_diagon
} and \cite{Choquet-Bruhat_09
}.
A different approach was proposed by Friedrich
\cite{friedrich98:_evolut
}, with the motivation to treat free initial boundary problems.
So he was able to write the relativistic Euler equations in Lagrangian
coordinates as a symmetric hyperbolic system by differentiating the
equations in an appropriate manner. 
This leads to a system with constraint equations, whose propagation
needs to be shown separately. 
The advantage of his system is the fact that it is more
suited to deal with initial free-boundary problems since in
Lagrangian coordinates the boundary is fixed.

Disconzi used Friedrich's approach to derive local existence and uniqueness
of classical solutions for the non--isentropic Einstein--Euler 
system~\cite{Disconzi_15},
using uniformly local Sobolev spaces, assuming the density to be strictly
positive and a smooth equation of state.
Another approach for the non--isentropic relativistic Euler equations
was presented by Walton \cite{Walton_arxiv_05
}, however, no local existence
and uniqueness system is known using this approach.

The purpose of these notes is to generalize our approach as provided
in \cite{BK8
} and present the non-isentropic relativistic Euler equations as a
symmetric hyperbolic system, which would enable us to prove similar
local existence and uniqueness theorem, therefore removing some of the
restrictions posed in the results of \cite{Disconzi_15}.

%% file: sec-02-non-isentropic.tex
\section{The relativistic Euler equations with entropy}
\label{sec:relativistic-case}
We now briefly introduce the notion of a relativistic perfect, but and
non-isentropic fluid. 
For more information and the thermodynamical background see for
example
\cite{Friedrich_Rendall_00
},
\cite{christodoulou95:_self
},
\cite{Choquet-Bruhat_09
}.
We consider the fluid in a prescribed Lorentzian manifold
$(\mathcal{M},g_{\alpha\beta})$, $\alpha,\beta=0,1,2,3$, and we chose units such
that the speed of light $c=1$. 
For a perfect fluid, the energy-momentum tensor takes the following
form
\begin{align}
  \label{E:EMTensorcDef}
  T^{\alpha \beta} = (\epsilon + p) u^{\alpha} u^{\beta}  + p g^{\alpha \beta}, 
\end{align}
where $\epsilon$ is the {proper energy density} of the fluid, $p$ is the
{pressure}, and $u^{\alpha}$ is the {four-velocity}, which is subject
to the normalization constraint
\begin{align}
  \label{E:uNormalizedcSquared}
  g_{\alpha \beta} u^{\alpha} u^{\beta} = -1. 
\end{align}
The Euler equations for a perfect fluid are (see e.g. \cite{christodoulou95:_self
})
\begin{align}
  \label{E:Euler} 
  \nabla_{\alpha} T^{\alpha \beta} &= 0 \qquad (\beta=0,1,2,3) \\
\label{E:nandulaw}
  \nabla_{\alpha}(n u^{\alpha}) &= 0,  
\end{align}
where $n$ is the \emph{proper number density} and $\nabla_\alpha$ denotes
the covariant derivative induced by the spacetime metric
$g_{\alpha\beta}$. 
As we will discuss in section \ref{sec:fluid-decomposition}, the
projection $u_{\beta}  \nabla_{\alpha} T^{\alpha \beta} = 0$ leads to the
energy equation 
\begin{equation}
\label{eq:sec-02-non-isentropic2}
  u^{\nu}\nabla_{\nu}\epsilon + (\epsilon+p) \nabla_{\nu}u^{\nu}    = 0.
\end{equation}

A non-isentropic fluid contains a thermodynamic variable $s$ that
represents the \emph{Entropy}, 
and satisfies the following thermodynamic relation, called Gibbs
relation,  \cite{Choquet-Bruhat_09
}
\begin{equation}
\label{eq:sec-02-non-isentropic1}
T ds= d \left( \frac{\epsilon}{n} \right)+ p d\left(  \frac{1}{n} \right),
\end{equation}
where $T$ denotes the temperature. 
As it was proven by Pichon
\cite{pichon65:_etude
}, the energy equation \eqref{eq:sec-02-non-isentropic2}, the
rest-mass conservation equation \eqref{E:nandulaw} and the Gibbs
relation \eqref{eq:sec-02-non-isentropic1} imply the following
relation for the entropy
\begin{equation}
\label{eq:sec-02-non-isentropic:1}
  u^{\alpha}\nabla_{\alpha}s = 0,
\end{equation}
which just expresses the fact that it is conserved along the fluid lines.

The equation of state specifies the relations between the number density 
$n$, entropy $s$, and the mass density $\epsilon$. We assume an equation 
of state is given by a nonnegative function
\begin{equation}
\label{eq:density}
 \epsilon =\epsilon(n,s), 
 \qquad n, s \geq 0.
\end{equation} 
 From laws of thermodynamics  (see e.g. 
\cite{Friedrich_Rendall_00}) it follows that the pressure is given by 
\begin{equation}
\label{E:pressure}
    p= n 
    \frac{\partial \epsilon}{\partial n} -\epsilon,
  \end{equation}
and the speed of sound is given by
 \begin{equation}
  \label{E:SpeedofSoundc}
    \sigma^2 {=}  \frac{\partial p}{\partial
    \epsilon}=  \frac{{\frac{\partial p}{ \partial 
n}}}{{\frac{ \partial \epsilon}{\partial n}}}.
  \end{equation}
  A fundamental  thermodynamic assumption is that the right--hand side of 
(\ref{E:SpeedofSoundc}) is positive, hence we require that
\begin{equation}
 \label{E:EOSAssumptions}
    \frac{\partial \epsilon}{\partial n}>0, \qquad
    \frac{\partial p}{\partial
    n} >0.
  \end{equation}
  
Another requirement is that $\sigma<1$, which means that the sound speed is 
always less than the speed of light.

\subsection{Energy conditions}

The  General Relativity literature  refers to three types of energy conditions 
(see e.g. \cite{Choquet-Bruhat_09}). The energy-momentum tensor 
$T^{\alpha\beta} $ satisfies:
\begin{enumerate}
 \item  The \emph{ weak energy condition},  if 
$T_{\alpha\beta}X^\alpha X^\beta\geq 0$ for all timelike vectors $X^\alpha$.

\item
The \emph{strong energy condition}, if
$[T_{\alpha \beta} - T g_{\alpha \beta}]X^{\alpha}X^{\beta} \geq 0$ for all 
timelike vectors $X^\alpha$, where $T=g_{\mu\nu}T^{\mu\nu}$.

\item
The \emph{dominant energy condition}, if 
    $-T^{\alpha}_{\ \beta} X^{\beta}$ is timelike  future-directed vector
    for all $X^\alpha $   future-directed timelike vector.
\end{enumerate}

Whenever $\epsilon\geq 0$ and $p\geq 0$, the perfect fluid satisfies the weak 
and strong energy conditions. If $\epsilon\geq p$, then it satisfies also the 
dominant energy condition,  see \cite{Choquet-Bruhat_09}. We shall see that the 
examples below meet all the  three energy conditions. 
    
\subsection{ Examples of an equation of state for the non--isentropic Euler 
equations}
\label{sec:equation-state-non}
A typical  non-isentropic equation of state is given by 
(see also
\cite{guo99:_format})
\begin{equation}
  \label{eq:sec-02-non-isentropic:3}
  \epsilon = n + \frac{A(s)}{\gamma-1}n^{\gamma},
\end{equation}
where $1<\gamma<2$ and $A(s)$ is a positive function. Equation 
(\ref{E:pressure}) implies that   $ p =  A(s)n^{\gamma}$, and from 
 \eqref{E:SpeedofSoundc} we can compute the speed of sound as follows,
\begin{equation}
\label{eq:sound:1}
\sigma^2=\frac{\gamma(\gamma-1)A(s)n^{\gamma-1}}{(\gamma-1)+\gamma 
A(s)n^{\gamma-1}}.
\end{equation} 
As a function of $n$, the speed of sound $\sigma $ is increasing and tends
to $\sqrt{\gamma-1} $ as $n$ tends to infinity. 
Hence the speed of sound is less than the speed of light. 
The equation of state \eqref{eq:sec-02-non-isentropic:3} also
satisfies the dominant energy condition, since
\begin{equation}
 \epsilon-p=n+\frac{(2-\gamma)A(s)n^{\gamma}}{\gamma-1}\geq0.
\end{equation}

Another example is a polytropic equation of state with index 
$\gamma=\frac{4}{3}$. We follow the convention of Choquet--Bruhat
\cite{Choquet-Bruhat_09
}, here
\begin{equation}
\label{eq:section-0002-non-isentropic:3}
p = \frac{K}{3}\left( \frac{3s}{4K} \right)^{\frac{4}{3}}n^{\frac{4}{3}} 
\qquad \text{and} \quad \epsilon=3p+n,
\end{equation}
where $K$ is a positive constant. We see that $ \frac{\partial 
\epsilon}{\partial n}=\frac{4K}{3}\left( \frac{3s}{4K} 
\right)^{\frac{4}{3}}n^{\frac{1}{3}}+1=\frac{p+\epsilon}{
n}$, hence (\ref{E:pressure}) is fulfilled. We also note that
\begin{equation*}
 p=n+K\left(\frac{3s}{4K} \right)^{\frac{4}{3}}n^{\frac{4}{3}},
\end{equation*}
and hence it is a particular case of the equation of state  
(\ref{eq:sec-02-non-isentropic:3}). So this equation of state also satisfies 
the dominant energy condition.

%% file: sec-03-system-p-u-S.tex
\section{The non--isentropic equations in symmetric hyperbolic form }
\label{sec:system-u=p-ualpha}

The equation of state (\ref{eq:density}) and the explicit formula of
the pressure (\ref{E:pressure}) allows us to express the pressure $p$
as a function of $n$ and $s$, which leads to consider
$U=(n,u^\alpha,s)$, $\alpha=0,1,2,3$ as the unknowns for the Euler
equations (\ref{E:Euler}) and (\ref{E:nandulaw}).

However, such an equation of state implies also that
$\nabla_{\alpha}p=\frac{\partial p}{\partial n}\nabla_{\alpha}n+
\frac{\partial p}{\partial s}\nabla_{\alpha}s$, which destroys the symmetry of the
corresponding matrices and makes it almost impossible to cast the Euler
equations in symmetric hyperbolic form.
The same problem occurs for the non-relativistic case, and there the
solution consists in using the pressure $p$ as a matter variable
instead of the density $n$.

That is why we take a similar approach here for the relativistic
equations and cast the equations in symmetric hyperbolic form.

Moreover, the resulting system is a more convenient starting point to
introduce the regularizing Makino variable.

\subsection{Symmetric Hyperbolic Systems}
\label{sec:init-value-probl}

We recall the definition of  symmetric hyperbolic
systems.
\begin{defn}[Symmetric hyperbolic system]
  \label{def:euler-rel:1}
  A first order quasi--linear $k\times k$ system is {\it symmetric
    hyperbolic system} in a region $G\subset \setR^k$, if it is of the form
  \begin{equation}
    \label{eq:publ-broken:7}
    L\lbrack   U \rbrack      =     
    A^\alpha (U)\partial_\alpha U  +
    B(U) = 0,
  \end{equation}
  where the matrices $A^\alpha(U) $ are symmetric and for every
  arbitrary $U\in G $, and there exists a covector $\xi_\alpha$ such that
  \begin{equation}
    \label{eq:publ-broken:8}
    \xi_\alpha   A^\alpha (U)
  \end{equation} is positive definite. 
  The covectors $\xi_\alpha$ for which (\ref{eq:publ-broken:8}) is
  positive definite, are called {\it spacelike with respect to
    equation }(\ref{eq:publ-broken:7}).
\end{defn}

\begin{rem}
  In most applications, and in particular, for initial value problems,
  it is essential that $A^0(U)$ is positive definite, and then system
  (\ref{eq:publ-broken:7}) takes the form
\begin{equation}
\label{eq:symm}
 A^0(U)\partial_t U=\sum_{i=1}^3A^i(U)\partial_{x^i} U+B(U).
\end{equation} 
To derive equation (\ref{eq:publ-broken:7}) in the above form requires
to show that $(1,0,0,0)$ is spacelike with respect to the equation.
Under the assumption that the speed of sound is less than 
one, we shall prove that the covector  $(1,0,0,0)$ belongs the future sound 
cone, and hence it is spacelike with respect to the equation 
(\ref{eq:publ-broken:7}).

\end{rem}

\subsection{Fluid decomposition}
\label{sec:fluid-decomposition}

First, we apply the well known fluid decomposition (see for example 
\cite{BK8
}) to equation (\ref{E:Euler}). 
We project $\nabla_\nu T^{\nu\beta}$ along the flow lines $u^\nu$, by 
$u_\beta\nabla_{\nu}T^{\nu\beta}$, and on the orthogonal subspace to the flow 
lines $\mathcal{O}$, by $ P_{\alpha\beta}\nabla_{\nu}T^{\nu\beta}$, where 
\begin{equation}
 P_{\alpha\beta}=g_{\alpha\beta}+u_\alpha u_\beta.
\end{equation} 
These projections result in
\begin{align}
  \label{eq:euler-rel:10}
  u^{\nu}\nabla_{\nu}\epsilon + (\epsilon+p) \nabla_{\nu}u^{\nu}    &= 0 \\
  \label{eq:non-isotropic-entropy-sh:26}
  (\epsilon+p)   P_{\alpha\beta}u^{\nu}\nabla_{\nu}{u^\beta} + P^\nu{}_\alpha    
 \nabla_\nu p     &=    0,
\end{align}
which together with the continuity equation \eqref{E:nandulaw} form a
system of equations. 
As we already pointed out the energy equation \eqref{eq:euler-rel:10},
together with the continuity equation \eqref{E:nandulaw} and the
thermodynamical relation \eqref{eq:sec-02-non-isentropic1} imply the
conservation of the entropy \eqref{eq:sec-02-non-isentropic:1}.
Moreover, we will also need that fact, that thanks to equation
\eqref{E:EOSAssumptions}, we can express $n$ as a function of $p$. 
All these considerations allow us to consider the following system of
equations:
\begin{align}
\label{eq:sec-03-system-p-u-S1}
  u^{\nu}\nabla_{\nu}n + n \nabla_{\nu}u^{\nu}    &= 0 \\
\label{eq:sec-03-system-p-u-S2}
  (\epsilon+p)   P_{\alpha\beta}u^{\nu}\nabla_{\nu}{u^\beta} + P^\nu{}_\alpha
  \nabla_\nu p     &=    0\\
\label{eq:sec-03-system-p-u-S3}
  u^{\alpha}\nabla_{\alpha}s &= 0.
\end{align}

\subsection{Modification of the fluid decomposed system}
\label{sec:modif-fluid-decomp}

In order to obtain a symmetric hyperbolic system we 
modify  the coupled 
equations (\ref{eq:sec-03-system-p-u-S1})-(\ref{eq:sec-03-system-p-u-S3}) 
 the following way. 
The normalisation condition (\ref{E:uNormalizedcSquared})
implies that
\begin{equation}
  \label{eq:rel-euler-symm:7}
  u_\beta u^\nu\nabla_\nu u^\beta= 0.
\end{equation}
So we add
\begin{math}
n u_\beta u^\nu \nabla_\nu u^\beta = 0
\end{math}
to equation (\ref{eq:sec-03-system-p-u-S1}),
\begin{math}
  u_\alpha u_\beta u^\nu\nabla_\nu u^\beta= 0
\end{math}
to (\ref{eq:sec-03-system-p-u-S2})
 and we obtain finally that
\begin{align}
  \label{eq:eineul:8}
  u^{\nu}\nabla_{\nu}n + n P^\nu{}_\beta\nabla_\nu u^\beta&= 0 \\
  \label{eq:publ-broken:3}
  (\epsilon+p)  \Gamma_{\alpha\beta}    u^{\nu}\nabla_{\nu}u^{\beta} +P^\nu{}_\alpha
  \nabla_\nu p  &= 0,
\end{align}
where
\begin{equation}
  \label{eq:rel-euler-symm:10}
  \Gamma_{\alpha\beta}= P_{\alpha\beta}+u_{\alpha}u_{\beta}= 
g_{\alpha\beta}+2u_{\alpha}u_{\beta}
\end{equation}
is a reflection with respect to the hyperplane $\mathcal{O}$.

We now use the equation of state (\ref{eq:density})
and (\ref{E:pressure}),  which allow us to express $p$ as 
a function of $n$ and $s$, that is, $p=p(n,s$).  Hence,
\begin{equation}
\label{eq:sec-03-system-p-u-S:3}
 \nabla_{\nu} p= \frac{\partial p}{\partial n}  \nabla_{\nu}n  +
 \frac{\partial p}{\partial s}  \nabla_{\nu}s,
\end{equation}
and  by the conservation of the entropy (\ref{eq:sec-02-non-isentropic:1}),  
we  conclude that
\begin{equation}
\label{eq:sec-03-system-p-u-S:4}
u^{\nu} \nabla_{\nu} p= \frac{\partial p}{\partial n}u^{\nu}
\nabla_{\nu} n + \frac{\partial p}{\partial s }u^{\nu}  
\nabla_{\nu}s  =\frac{\partial p}{\partial n}u^{\nu}
\nabla_{\nu}n.
\end{equation}
So we finally obtain the system
\begin{align}
\label{eq:non-isotropic-entropy-sh:113}
  u^{\nu}\nabla_{\nu} p + n\frac{\partial p }{\partial n} 
P^\nu{}_\beta\nabla_\nu u^\beta&= 0 \\
\label{eq:non-isotropic-entropy-sh:114}
  (\epsilon + p)  \Gamma_{\alpha\beta}    u^{\nu}\nabla_{\nu}u^{\beta} +P^\nu{}_\alpha  \nabla_\nu p  &= 0\\
\label{eq:non-isotropic-entropy-sh:115}
  u^{\alpha}\nabla_{\alpha}s&=0.
\end{align}

\begin{rem}[The pressure as a matter variable]
  The idea of using the pressure as a matter variable instead of the
  density is widely used in the non-relativistic case, see for example
  \cite{smoller83:_shock_waves_react
  }. 
  In the relativistic case, Guo and
  Tahvildar-Zadeh \cite{guo99:_format
  } presented the following system for the variables $(p,u^\alpha,s)$
  \begin{align}
    \label{eq:non-isotropic-entropy-sh:72}
    \displaystyle\frac{1}{(\epsilon+p)\sigma}u^\nu\partial_\nu p + \sigma \partial_\nu
    u^\nu
    & =  0 \\
    \label{eq:non-isotropic-entropy-sh:73}
    \sigma P^{\mu\nu}\partial_\nu p + (\epsilon+p)\sigma u^\nu \partial_\nu u^\mu  & =  0\\
    \label{eq:non-isotropic-entropy-sh:74}
    u^\nu \partial_\nu s & = 0.
  \end{align}
  It should be pointed out, that this system, however, is not symmetric
  hyperbolic as it can be easily checked. 
\end{rem}

\subsection{Symmetric hyperbolic form}
\label{sec:symm-hyperb-form}

We now write system
(\ref{eq:non-isotropic-entropy-sh:113})-(\ref{eq:non-isotropic-entropy-sh:115})
in matrix form
\begin{equation}
\label{eq:sec-03-system-p-u-S:6}
  \begin{pmatrix}
    u^{\nu} &  n \frac{\partial p}{\partial n} P^{\nu}{}_{\beta} & 0 \\
    P^{\nu}{}_{\alpha}       & (\epsilon +p)\Gamma_{\alpha\beta} u^{\nu}       & 0 \\
    0                      & 0                                                   & u^{\nu}
  \end{pmatrix}
  \nabla_{\nu}
  \begin{pmatrix}
    p                                                                                \\
    u^{\alpha}                                                                        \\
   s
  \end{pmatrix}
  = 0.
\end{equation}
These matrices are not symmetric, but they can be cast into a
symmetric form by choosing an appropriate multiplier, for example, we
multiply the second row of the matrices by
\begin{math}
 n\frac{\partial p}{\partial n}
\end{math},
and then  we obtain 
\begin{equation}
\label{eq:sec-03-system-p-u-S:8}
  \begin{pmatrix}
    u^{\nu} & n\frac{\partial p}{\partial n} P^{\nu}{}_{\beta}                  
            & 0 \\
n\frac{\partial p}{\partial n }    P^{\nu}{}_{\alpha} & 
n \frac{\partial p}{\partial n}(\epsilon +p)\Gamma_{\alpha\beta} u^{\nu} 
& 0 \\
    0                               & 0                                                                                & u^{\nu}
  \end{pmatrix}
  \nabla_{\nu}
  \begin{pmatrix}
    p                                                                                \\
    u^{\alpha}                                                                        \\
    s
  \end{pmatrix}
  = 0,
\end{equation}
which are symmetric matrices. 

In fact, it turns out that system \eqref{eq:sec-03-system-p-u-S:8} is
a symmetric hyperbolic system. 
The following theorem gives a precise statement.
\begin{thm}
  \label{thm:2}
  Let $\epsilon$ in (\ref{eq:density}) be nonnegative density function, the
  pressure $p$ be defined by (\ref{E:pressure}) and assume conditions
  (\ref{E:EOSAssumptions}). 
  Then the Euler equations (\ref{E:Euler})-(\ref{E:nandulaw}) coupled
  with the constraint (\ref{E:uNormalizedcSquared})
  can be written as a symmetric hyperbolic system. 
  Moreover, under the assumption that the speed of sound is less than
  the speed of light, the matrix $A^0$ is positive definite and
  therefore the Euler equations (\ref{E:Euler})-(\ref{E:nandulaw})
  form are symmetric hyperbolic system as specified in equation
  (\ref{eq:symm}).
\end{thm}

\begin{proof}

To show that the system (\ref{eq:sec-03-system-p-u-S:8}) is symmetric 
hyperbolic we need to show that $\xi_\alpha A^\alpha(U)$ is positive 
definite for some covectors $\xi_\alpha$.
For that we slightly rewrite system (\ref{eq:sec-03-system-p-u-S:8}). Using 
equations (\ref{E:SpeedofSoundc}) and   (\ref{E:SpeedofSoundc}) we see that
\begin{equation}
 n \frac{\partial p}{\partial n}=  \frac{\partial p}{\partial 
\epsilon} n\frac{\partial \epsilon}{\partial n}
  =\sigma^2 (\epsilon+p),
\end{equation}
hence (\ref{eq:sec-03-system-p-u-S:8}) is  equivalent to system
\begin{equation}
\label{eq:sec-03-system-p-u-S:7}
  \begin{pmatrix}
    u^{\nu}                          & \sigma^2 \left( \epsilon+p \right) P^{\nu}{}_{\beta}                              & 0 \\
\sigma^2 \left( \epsilon+p \right)    P^{\nu}{}_{\alpha} & 
\sigma^2 \left( \epsilon+p \right)^{2}\Gamma_{\alpha\beta} u^{\nu} & 0 \\
    0                               & 0                                                                                & u^{\nu}
  \end{pmatrix}
  \nabla_{\nu}
  \begin{pmatrix}
    p \\
    u^{\alpha}\\
   s
  \end{pmatrix}
  = 0.
\end{equation}

Now we compute the principal symbol of system
(\ref{eq:sec-03-system-p-u-S:7}).
For each $\xi_\alpha\in T_x^*V$ the principal symbol is a linear map from
$\mathbb{R}\times E_x$ to $\mathbb{R}\times F_x$, where $E_x$ is a fiber in
$T_xV$ and $F_x$ is a fiber in the cotangent space $T_x^*V$.
In local coordinates $\nabla_{\nu} =\partial_{\nu} + \Gamma$, where
$\Gamma=\Gamma (g^{\gamma\delta},\partial g_{\alpha\beta})$ denotes the
Christoffel symbols, hence the principal symbol of system
(\ref{eq:sec-03-system-p-u-S:7}) is
\begin{equation}
  \label{eq:Initial:4}
\xi_\nu A^\nu= \left(
    \begin{array}{c|ccc|c}
        (u^\nu \xi_\nu)   &  & \sigma^2 \left( p+\epsilon \right) P^{\nu}{}_{\beta}\xi_\nu  &   & 0  \\ \hline
      &  &                     &  &  \\
      \sigma^2 \left( p+\epsilon \right)  P^{\nu}{}_{\alpha}\xi_\nu
      &  & \sigma^2 \left( p+\epsilon \right)(u^\nu\xi_\nu)\Gamma_{\alpha\beta}        &    & 0\\ 
      &  &                     &  &  \\\hline
    0  &  &            0        &  &  (u^\nu\xi_\nu)\\
    \end{array}
  \right).
\end{equation}
The characteristics are the set of covectors $\xi_\nu$ for which $(\xi_\nu
A^\nu)$ is not an isomorphism. Hence the characteristics are the zeros
of 
\begin{equation}
 Q(\xi) \eqdef \det(\xi_\nu A^\nu).
\end{equation}

The geometric advantages of fluid decomposition are the following.
The operators in the blocks of the matrix (\ref{eq:Initial:4}) are
the projection $P^{\nu}{}_{\alpha}$, on the hyperplane $\mathcal{O}$ that is 
orthogonal to the flow lines,  and the reflection  $ \Gamma_{\alpha\beta}$, 
with respect to the same hyperplane. Therefore, the following relations hold:
\begin{displaymath}
  \Gamma^{\alpha\gamma}\Gamma_{\gamma\beta}=\delta_\beta{}^\alpha,
  \qquad \Gamma^{\alpha\gamma}P_\gamma{}^\nu=P^{\alpha\nu}\qquad
  \text{and}\qquad P_\beta{}^\alpha P_{\alpha}{}^\nu=P^{\nu}{}_\beta,
\end{displaymath}
which yields
\begin{equation}
  \label{eq:Initial:5}
 \begin{split}
  & \left(
    \begin{array}{c|ccc|c}
      1 &  & 0                     &  & 0 \\ \hline
        &  &                       &  &   \\
      0 &  & \Gamma^{\alpha\gamma} &  & 0 \\
        &  &                       &  &   \\ \hline
      0 &  & 0                     &  & 1
    \end{array}
  \right)
  \left(\xi_\nu A^\nu\right) \\ = & \left(
    \begin{array}{c|ccc|c}
(u^\nu \xi_\nu)                                              &  & \sigma^{2}\left( p+\epsilon \right)      P^{\nu}{}_{\beta}\xi_\nu                        &  & 0               \\ \hline
                                                             &  &                                                                                          &  &                 \\
\sigma^{2}\left( p+\epsilon \right)   P^{\alpha\nu}{}\xi_\nu &  & \sigma^{2}\left( p+\epsilon \right)  (u^\nu\xi_\nu)\left(\delta^{\alpha}_{\beta} \right) &  & 0               \\\hline
0                                                            &  & 0                                                                                        &  & (u^\nu \xi_\nu) \\
    \end{array}
  \right).
 \end{split}
\end{equation}

It is now fairly easy to calculate the determinant of the right-hand
side of (\ref{eq:Initial:5}) and we have
\begin{dmath*}
  \det
\left(
    \begin{array}{c|ccc|c}
(u^\nu \xi_\nu)                                 &  & \sigma^{2}\left( p+\epsilon \right)      P^{\nu}{}_{\beta}\xi_\nu                       &  & 0             \\ \hline
                                              &  &                                                                                     &  &               \\
\sigma^{2}\left( p+\epsilon \right)   P^{\alpha\nu}{}\xi_\nu &  & \sigma^{2}\left( p+\epsilon \right)  (u^\nu\xi_\nu)\left(\delta^{\alpha}_{\beta} \right) &  & 0             \\\hline
0                                             &  & 0                                                                                   &  & (u^\nu \xi_\nu) \\
    \end{array}
  \right)\\
  =\sigma^2 \left( p+\epsilon \right)^{2} 
(u^\nu\xi_\nu)^4\left\{(u^\nu\xi_\nu)^2 -\sigma^2    P^{\alpha\nu}\xi_\nu 
P_\alpha ^\nu\xi_\nu\right\}.
\end{dmath*}
Since $P_\beta^\alpha$ is a projection,
\begin{equation*}
  P^{\alpha\nu}\xi_\nu P_\alpha ^\nu\xi_\nu= g^{\nu\beta}\xi_\nu
  P^{\alpha}_\beta P_\alpha ^\nu\xi_\nu
  =g^{\nu\beta}\xi_\nu P^{\nu}{}_\beta \xi_\nu
  =P^{\nu}{}_\beta\xi_\nu  \xi^\beta,
\end{equation*}
and since $\Gamma_\beta^\gamma$ is a
reflection,
\begin{equation}
\label{eq:Initial:13}
  \det\left(
    \begin{array}{c|cc|c}
      1 &  & 0                     & 0 \\ \hline
      0 &  & \Gamma^{\alpha\gamma} & 0 \\\hline
      0 &  & 0                     & 1
    \end{array}
  \right) =\det\left(g^{\alpha\beta}\Gamma_\beta^\gamma\right)=
 -\left({\rm
      det}\left(g_{\alpha\beta}\right)\right)^{-1}>0.
\end{equation}
Consequently,
\begin{equation}
  \label{eq:Initial:11}
  Q(\xi )=  \det (\xi_\nu A^\nu  ) =- \sigma^2 \left( p+\epsilon \right)^{2}
  \det(g_{\alpha\beta})  ( u^\nu\xi_\nu)^4
  \left\{ ( u^\nu\xi_\nu)^2 - \sigma^2
    P^{\alpha}{}_{\beta}\xi_\alpha \xi^\beta \right\}
\end{equation}
and therefore the characteristic covectors are given by two simple
equations:
\begin{eqnarray}
  \label{eq:publ-broken:39}
  \xi_\nu u^\nu & = &0 \\
  \label{eq:publ-broken:40} (\xi_\nu u^\nu)^2 - \sigma^2
  P^{\alpha}{}_{\beta}\xi_\alpha \xi^\beta & = &0.
\end{eqnarray}
\begin{rem}
  \label{rem:euler-rel:7}
  The characteristics conormal cone is a union of two
  hypersurfaces in $T_x^*V$.  One of these hypersurfaces is given by
  the condition (\ref{eq:publ-broken:39}) and it is a three
  dimensional hyperplane $\mathcal{O}$ with the normal $u^\alpha$.
  The other hypersurface is given by the condition
  (\ref{eq:publ-broken:40}) and forms a three--dimensional cone, the 
  so--called, {\it sound cone}.
\end{rem}

Let us now consider the timelike vector $u_{\nu }$ and  insert the
covector  $-u_\nu$ into the principal symbol
(\ref{eq:Initial:4}), then
\begin{equation*}
  -u_\nu A^\nu=\left(
    \begin{array}{c|ccc|c}
      1 &  & 0                    &  &0    \\
      \hline
        &  &                      &  &   \\
      0 &  & \sigma^2(p+\epsilon)\Gamma_{\alpha\beta} &  & 0 \\
        &  &                      &  &   \\\hline
      0 &  & 0                    &  & 1 \\
    \end{array}
  \right) \label{kap3.mk3}
\end{equation*}
is a positive definite matrix . 
Indeed, $\Gamma_{\alpha\beta}$ is a reflection with respect to a
hyperplane having a timelike normal, and as in (\ref{eq:Initial:13}) we see 
that $\det(\Gamma_{\alpha\beta})>0$. 
Hence, $-u_\nu$ is a spacelike covector with respect to the
hydrodynamical equations (\ref{eq:sec-03-system-p-u-S:7}).
Herewith, we have shown relatively elegant and elementary that the
relativistic hydrodynamical equations are symmetric hyperbolic.

We want now to show that $A^0$ is positive definite. To do that it suffices to 
show  that the covector $\zeta_{\nu}=(1,0,0,0)$
is also spacelike with respect to the system (\ref{eq:sec-03-system-p-u-S:7}). Since
$P^\alpha{}_{\beta}u_\alpha=0$, the covector $-u_\nu$ belongs to the
sound cone
\begin{equation}
  \label{eq:Initial:9}
  (\xi_\nu u^\nu)^2 - \sigma^2P^{\alpha}{}_{\beta}\xi_\alpha
  \xi^\beta>0.
\end{equation}
Inserting $\zeta_\nu=(1,0,0,0)$ the right-hand side of
(\ref{eq:Initial:9}), yields
\begin{equation}
  \label{eq:Initial:10}
  (u^0)^2(1-\sigma^2)-\sigma^2 g^{00}.
\end{equation}
Under the assumption sound velocity is less than the speed of light,
that is $\sigma^2=\frac{\partial p}{\partial \epsilon}<c^2=1$, we conclude that
(\ref{eq:Initial:10}) is positive, and hence $\zeta_\nu=(1,0,0,0)$ also
belongs to the sound cone (\ref{eq:Initial:9}). 
Hence, the vector $-u_\nu$ can be continuously deformed to $\zeta_\nu$
while condition (\ref{eq:Initial:9}) holds along the deformation path. 
Consequently, the determinant of (\ref{eq:Initial:11}) remains
positive under this process and hence $\zeta_{\nu}A^\nu=A^{0}$ is also
positive definite.
\end{proof}

%% file: sec-04-makino.tex
\section{Symmetrization and regularization}

In the case of a physical vacuum, that is, if the density or the
pressure vanish in certain regions, or fall-off at infinity, the
symmetrization we obtained in Section \ref{sec:system-u=p-ualpha}
breaks down. 
The reason for this can be seen easily by inspecting the matrix
$A^{0}(U)$ which is no longer uniformly positive definite if the
pressure approaches zero. 
Makino symmetrised and regularised the Euler-Poisson system by
introducing a new nonlinear matter variable $w=M(\rho)$
\cite{makino_86
}, so that the matrix $A^{0}(U)$ remains uniformly positive even for
$\rho=0$.
Later Makino generalised his regularisation to the non isentropic
Euler-Poisson system
\cite{makino87:_sur_solut_local
}, starting with a system for $(p,u^{\alpha},s)$. 
We follow this strategy but, naturally, have to modify it due to the
more complicated character of our equations.

So, we start with system
\eqref{eq:non-isotropic-entropy-sh:113}--\eqref{eq:non-isotropic-entropy-sh:115}
\begin{align}
\label{eq:sec-04-makino2}
  u^{\nu}\nabla_{\nu} p + n\frac{\partial p }{\partial n} 
P^\nu{}_\beta\nabla_\nu u^\beta&= 0 \\
\label{eq:sec-04-makino3}
  (\epsilon + p)  \Gamma_{\alpha\beta}    u^{\nu}\nabla_{\nu}u^{\beta} +P^\nu{}_\alpha  \nabla_\nu p  &= 0\\
  u^{\alpha}\nabla_{\alpha}s&=0.
\end{align}
and replace $p$ by $w=w(p)$. 
Then we multiply equation \eqref{eq:sec-04-makino2} by
$\kappa^2(w,s)\frac{\partial w}{\partial p}$ where $\kappa$ is a positive function we specify
later in order to simplify our calculations. 
Moreover, we divide equation \eqref{eq:sec-04-makino3} by
$(\epsilon+p)$, then equations \eqref{eq:sec-04-makino2} and
\eqref{eq:sec-04-makino3} written in matrix form, take the following
form
\begin{equation}
  \label{eq:makino:11}
  \begin{pmatrix}
 \kappa^{2}   u^{\nu}& \kappa^{2}n\frac{\partial p}{\partial n}\frac{\partial w}{\partial p} P^{\nu}{}_{\beta} & 0
    \\
    \frac{1}{\left( \epsilon+p \right)}\frac{\partial p}{\partial w}
    P^{\nu}{}_{\alpha} &
    \Gamma_{\alpha\beta}u^{\nu}
    & 0 \\
    0 & 0 & u^{\nu}
  \end{pmatrix}
  \nabla_{\nu}
  \begin{pmatrix}
    w\\
    u^{\alpha}\\
    s
  \end{pmatrix}
  = 0,
\end{equation}
The matrices \eqref{eq:makino:11} are symmetric provided that 
\begin{equation}
  \label{eq:sec-04-makino1}
  \kappa^2n \frac{\partial w}{\partial n}
=  \kappa^2n \frac{\partial w}{\partial p}\frac{\partial p}{\partial n}
=\frac{1}{\epsilon+p}\frac{\partial p}{\partial w},
\end{equation}
which results in 
\begin{equation}
\label{eq:sec-04-makino8}
w = \int\limits_{}^{}\frac{1}{\kappa}\left( \frac{1}{\left(\epsilon+p \right)n}
\right)^{\frac{1}{2}} \left( \frac{\partial n}{\partial p} 
\right)^{\frac{1}{2}}dp.
\end{equation}
We will now, in the subsection below, calculate an explicit form of
this new variable using the equation of state
\eqref{eq:sec-02-non-isentropic:3} presented in section
\ref{sec:equation-state-non}.

\subsection{The Makino variable for the equation of state 
(\ref{eq:sec-02-non-isentropic:3})}

For this equation of state we easily compute 
\begin{align}
  \label{eq:sec-04-makino4}
  \epsilon+p &= n + 
\frac{1}{\gamma-1}A(s)n^{\gamma}+p=n+\frac{\gamma}{\gamma-1}p,\\
 n \frac{\partial p}{\partial n}&=\gamma p
\end{align}
and 
\begin{equation}
 n=A^{-\frac{1}{\gamma}}(s)p^{\frac{1}{\gamma}}.
\end{equation} 
This allows us to calculate 
\begin{align*}
  \frac{1}{\left( \epsilon+p \right)n \frac{\partial p}{\partial n}} &= 
\frac{1}{\left( n+\frac{\gamma}{\gamma-1}p 
\right)p\gamma}=\frac{1}{\gamma}\frac{1}{np+\frac{\gamma}{\gamma-1}p^2}\\
&=\frac{1}{\gamma}  
\frac{1}{A^{-\frac{1}{\gamma}}(s)p^{1+\frac{1}{\gamma}}+\frac{\gamma}{\gamma-1}
p^2 } =\frac { 1 } { \gamma} \left( 
\frac{1}{A^{-\frac{1}{\gamma}}(s)+\frac{\gamma}{\gamma-1}p^{1-\frac{1}{\gamma}}}
 \right)\frac{1}{p^{1+\frac{1}{\gamma}}}.
\end{align*}
Keeping in mind the symmetry condition \eqref{eq:sec-04-makino1}, we
see that setting
\begin{equation}
\label{eq:sec-04-makino6}
\kappa^2 = \left( \left( \frac{2\gamma}{\gamma-1} \right)^2 
\frac{1}{\gamma}\frac{1}{A^{-\frac{1}{\gamma}}(s)+\frac{\gamma}{\gamma-1}p^{
\frac { \gamma-1 } { \gamma }}} \right),
\end{equation}
implies that $\frac{\partial w}{\partial 
p}=\frac{\gamma-1}{2}p^{-\frac{\gamma-1}{2\gamma}}$,
which leads to 
\begin{equation}
\label{eq:sec-04-makino7}
w= p^{\frac{\gamma-1}{2\gamma}}
\end{equation}
and 
\begin{equation}
\label{eq:sec-04-makino8a}
\kappa^2 (w,s)= \left( \left( \frac{2\gamma}{\gamma-1} \right)^2 
\frac{1}{\gamma}\frac{1}{A^{-\frac{1}{\gamma}}(s)+\frac{\gamma}{\gamma-1}w^2} 
\right).
\end{equation}

So we conclude the Euler equations (\ref{E:Euler})-(\ref{E:nandulaw})
coupled with the constraint (\ref{E:uNormalizedcSquared}) can be
written in the form
\begin{equation}
  \label{eq:makino:12}
  \begin{pmatrix}
 \kappa^{2}   u^{\nu}& \kappa^{2}\frac{\gamma(\gamma-1)}{2}w P^{\nu}{}_{\beta} & 
0
    \\
    \kappa^{2}\frac{\gamma(\gamma-1)}{2}w
    P^{\nu}{}_{\alpha} &
    \Gamma_{\alpha\beta}u^{\nu}
    & 0 \\
    0 & 0 & u^{\nu}
  \end{pmatrix}
  \nabla_{\nu}
  \begin{pmatrix}
    w\\
    u^{\alpha}\\
    s
  \end{pmatrix}
  = 0,
\end{equation}
which is symmetric and regular when $p$, or equivalently $w$ approaches zero.